\documentclass[twocolumn,superscriptaddress,floatfix,preprintnumbers,letterpaper]{revtex4-1}
\usepackage{longtable}
\usepackage{array}
\usepackage{dcolumn}
\usepackage{bm}
\usepackage[colorlinks, pdfborder={0 0 0}, plainpages=false]{hyperref}
\usepackage{graphicx}
\usepackage{xspace}
\usepackage[usenames,dvipsnames]{color}
\usepackage{amssymb}
\usepackage[normalem]{ulem} 
\usepackage{lineno}
\usepackage{letltxmacro}
\usepackage{amsfonts}
\usepackage{amsmath}
\usepackage{amssymb}
\usepackage{bm}
\usepackage{dcolumn}
\usepackage{epsfig}
\usepackage{graphicx}
\usepackage{graphics}
\usepackage[latin1]{inputenc}
\usepackage{latexsym}
\usepackage{rotating}
\usepackage{hyperref}
\usepackage[usenames]{color}
\usepackage{float}
\usepackage{xspace} 
\usepackage{mathrsfs}
\usepackage{subfigure}
\usepackage{multirow}
\usepackage{booktabs}
\usepackage{soul}

\usepackage{colortbl}
\usepackage{etoolbox}
\usepackage[table]{xcolor}

\def\prd{{\it Phys. Rev.} D~}

\def\apjl{{\it Astrophys. J.} Lett~}
\def\apj{{\it Astrophys. J.}}

\def\PRD{{\it Phys. Rev.} D~}

\def\mnras{{\it MNRAS~}}

\def\lesssim{\mathrel{\hbox{\rlap{\hbox{\lower4pt\hbox{$\sim$}}}\hbox{$<$}}}}
\def\gtrsim{\mathrel{\hbox{\rlap{\hbox{\lower4pt\hbox{$\sim$}}}\hbox{$>$}}}}
\def\alt{\mathrel{\hbox{\rlap{\hbox{\lower4pt\hbox{$\sim$}}}\hbox{$<$}}}}
\def\agt{\mathrel{\hbox{\rlap{\hbox{\lower4pt\hbox{$\sim$}}}\hbox{$>$}}}}

\usepackage{mathtools}
\DeclarePairedDelimiter\abs{\lvert}{\rvert}

\def\gta{\ifmmode {\mathbin{\lower 3pt\hbox   
    {$\,\rlap{\raise 5pt\hbox{$\char'076$}}\mathchar"7218\,$}}}
    \else {${\mathbin{\lower 3pt\hbox
    {$\rlap{\raise 5pt\hbox{$\char'076$}}\mathchar"7218\,$}}}
    $}\fi}
\def\lta{\ifmmode {\,\mathbin{\lower 3pt\hbox   
    {$\,\rlap{\raise 5pt\hbox{$\char'074$}}\mathchar"7218\,$}}}
    \else {${\mathbin{\lower 3pt\hbox
    {$\rlap{\raise 5pt\hbox{$\char'074$}}\mathchar"7218\,$}}}
    $}\fi}

\newcommand{\beq}{\begin{equation}}
\newcommand{\eeq}{\end{equation}}
\newcommand{\bea}{\begin{eqnarray}}
\newcommand{\eea}{\end{eqnarray}}

\definecolor{darkperiwinkle}{RGB}{102, 102, 128}

\newcommand{\NCSA}{\affiliation{NCSA, University of Illinois at Urbana-Champaign, Urbana, Illinois 61801, USA}}
\newcommand{\ANCSA}{\affiliation{Department of Astronomy, University of Illinois at Urbana-Champaign, Urbana, Illinois 61801, USA}}
\newcommand{\PNCSA}{\affiliation{Department of Physics, University of Illinois at Urbana-Champaign, Urbana, Illinois 61801, USA}}
 
\newcommand{\ECE}{\affiliation{Department of Electrical and Computer Engineering, University of Illinois at Urbana-Champaign, Urbana, Illinois 61801, USA}}

\newcommand{\UNI}{\affiliation{The University of Illinois Laboratory High School, University of Illinois at Urbana-Champaign, Urbana, Illinois 61801, USA}}

\newcommand{\AR}{\affiliation{Department of Physics, University of Arizona, Tucson, Arizona 85721, USA}}
\newcommand{\STAN}{\affiliation{Institute for Computational and Mathematical Engineering, Stanford University, Stanford, California 94305, USA}}
\newcommand{\MS}{\affiliation{Microsoft, One Microsoft Way, Redmond, Washington 98052, USA}}
\newcommand{\OR}{\affiliation{Oracle Corporation, Seattle, Washington 98101, USA}}

\definecolor{light-gray}{gray}{0.9}

\begin{document}

\title{Physics of eccentric binary black hole mergers: A numerical relativity perspective}

\author{E. A. Huerta}\NCSA\ANCSA
\author{Roland Haas}\NCSA
\author{Sarah Habib}\NCSA\PNCSA
\author{Anushri Gupta}\NCSA\ANCSA
\author{Adam Rebei}\NCSA\UNI
\author{Vishnu Chavva}\NCSA\PNCSA
\author{Daniel Johnson}\NCSA\STAN
\author{Shawn Rosofsky}\NCSA\PNCSA
\author{Erik Wessel}\NCSA\AR
\author{Bhanu Agarwal}\NCSA\MS
\author{Diyu Luo}\NCSA\OR
\author{Wei Ren}\NCSA\ECE

\date{\today}

\begin{abstract}
\noindent Gravitational wave observations of eccentric binary black hole mergers will provide unequivocal evidence for the formation of these systems through dynamical assembly in dense stellar environments. The study of these astrophysically motivated sources is timely in view of electromagnetic observations, consistent with the existence of stellar mass black holes in the globular cluster M22 and in the Galactic center, and the proven detection capabilities of ground-based gravitational wave detectors. In order to get insights into the physics of these objects in the dynamical, strong-field gravity regime, we present a catalog of 89 numerical relativity waveforms that describe binary systems of non-spinning black holes with mass-ratios \(1\leq q \leq 10\), and initial eccentricities as high as \(e_0=0.18\) fifteen cycles before merger. We use this catalog to quantify the loss of energy and angular momentum through gravitational radiation, and the astrophysical properties of the black hole remnant, including its final mass and spin, and recoil velocity. We discuss the implications of these results for gravitational wave source modeling, and the design of algorithms to search for and identify eccentric binary black hole mergers in realistic detection scenarios.  
\end{abstract}

\pacs{Valid PACS appear here}

\maketitle

\section{Introduction}
\label{intro}

The gravitational wave (GW) detection of several binary black hole (BBH) mergers~\cite{DI:2016,secondBBH:2016,thirddetection,fourth:2017,GW170608,o1o2catalog}, 
and the first multi-messenger observation of two colliding neutron stars (NSs) in gravitational and electromagnetic waves~\cite{bnsdet:2017}
have shed light into the nature of gravity in the most extreme astrophysical settings, and has 
unveiled the identity of the central engines that power the most energetic electromagnetic explosions in the Universe~\cite{mma:2017arXiv,2017Sci...358.1556C,kiloGW170817:2017,grb:2017ApJ}, 
while also providing the means to put at work visionary methods to use GWs to quantify the rate of 
expansion of the Universe~\cite{GWH:NaturA,Maya:2018F,Schutz:1986Nature,HoHu:2005ApJH}. 

Along this trail of discovery, it has also become evident that numerical relativity (NR) 
plays a central role to understand the physics of GW sources, and to inform the development 
of signal-processing algorithms to detect and characterize these astrophysical events~\cite{Chu:2016CQG,Mroue:2013,Kumar:2016dhh,NRI:2016,NRGW1509142,Lange:2017PhRvD}, and 
astrophysical sources that still await discovery~\cite{geodf:2017a,geodf:2017b,hshen:2017,shen_icassp,shengeorge:PhysRevD97,huerta:2018PhRvD,Adam:2018arXiv,Hinder:2010,Huerta:2017a,sper:2008PhRvDS,hinder:2017a,sum:2009CQGra,kotake:2013CRPhy,Tanja:2018a,Fou:2018arX,rad:2018arXiv18}.

In preparation for the characterization of BBH mergers whose astrophysical properties 
span a parameter space that has not yet been probed by existing GW detections, several 
NR groups are working in earnest to construct large-scale NR waveform catalogs~\cite{Mroue:2013,RITcatalog:2017,GaTechcatalog:2016}. Since these activities have thus far focused on the study of quasi-circular BBH mergers, 
in this article we fill in a critical void in the literature by presenting a comprehensive study of the physics of moderately eccentric BBH mergers. 

The rationale for this study is multifold. From the perspective of electromagnetic 
observations, recent findings are consistent with the existence of stellar-mass BHs 
in the vicinity of the Galactic center, and in the Galactic Cluster M22~\cite{galcen:2018,Sippel:2013,Strader:2012,ssm:2018}. These observations have triggered the development of numerical models that provide a realistic 
description of the formation
and retention of BBHs in dense stellar environments, correcting previous calculations based on \(N\)-body simulations that did not include post-Newtonian corrections~\cite{Blanchet:2006} to model the orbital dynamics of these systems, thereby underestimating the merger rates of these systems by orders of magnitude~\cite{sam:2017ApJ,Samsing:2014,sam:2018PhRvD3014S,Leigh:2018MNRAS,ssm:2017,ssm:2018,samsing:2018ApJ140S,lisa:2018b,Huerta:2009,sam:2017ApJ84636S,sam:2018MNRAS1548S,Huerta:2015a,sam:2019MNRAS30S,sam:2018PhRvD3014S,sam:2018MNRAS5436S,Huerta:2014,Anton:2014,samdor:2018MNRAS5445S,samdorII:2018MNRAS4775D,samdor:2018arXiv64S,samjoh:2018Z,rocarl:2018PhRvDR,kremerjoh:2018aK,lopez:2018L,hoang:2017APJ,gon:2017,hpoang:2017,lisa:2018a,MikKoc:2012,Naoz:2013,gondkoc:2018G,antonras:2016ApJ7A,Huerta:2013a,arcakoc:2018A,takkoc:2018ApJT,gondkoc:2018ApJ5G,antoni:2018A,Anto:2015arXiv}. In summary, we have evidence for the existence of stellar-mass BHs that may form eccentric compact binary systems in dense stellar environments, and consequently be detected through GW emission. Through this study, we provide new insights into the physics of these astrophysically motivated sources.

Furthermore, as discussed in~\cite{Sergey:2016}, no matched-filtering algorithm 
has been presented in the literature that is tailored for the detection of 
\(\ell=\abs{m}=2\) eccentric waveforms~\cite{Sergey:2016}. However, signal processing algorithms based on deep neural networks have been used to demonstrate that moderately eccentric BBH mergers can be detected and characterized from real LIGO noise, considering both NR waveforms that only include the leading order quadrupole mode \(\ell=\abs{m}=2\)~\cite{geodf:2017a,geodf:2017b,hshen:2017,shengeorge:PhysRevD97}, and higher-order waveform multipoles~\cite{Adam:2018arXiv}. We expect that this NR waveform catalog may be used to quantify the sensitivity of burst searches, and of next-generation  neural network models that are tailored to 
detect and characterize eccentric BBH mergers. In summary, activities around modeling, detection and 
characterization of eccentric BBHs are reaching the required level of maturity to establish or rule out the existence of
compact binary populations in dense stellar environments. 

To advance our understanding of the physics of compact binary populations in dense stellar environments, in this article we introduce a NR waveform catalog that 
describes eccentric BBH mergers, and utilize it to get insights into the dynamics of these GW sources, e.g., the energy and angular momentum loss through GW emission, and the astrophysical properties of the BH remnant, i.e., its final mass and spin as a function of initial eccentricity and mass-ratio, as well as the recoil velocity of the BH remnant. These studies will 
inform ongoing GW modeling efforts, and the development of signal-processing algorithms to search for
and identify these sources. This article is organized as follows. Section~\ref{cat} describes the properties of our 
NR catalog. In Section~\ref{enbbh} we compute the energy and angular momentum radiated away through GW emission, making pair-wise comparisons between NR waveforms that include the \(\ell=\abs{m}=2\) mode or higher-order waveform multipoles. In Section~\ref{fin} we compute the astrophysical properties of the BH remnants, and compare these results with those obtained for quasi-circular BBH mergers. We describe the relevance of these analyses in terms of GW modeling efforts for eccentric BBH mergers in Section~\ref{mol}. We summarize our findings and outline future directions of work in Section~\ref{end}. 


\section{Numerical relativity catalog}
\label{cat}

\begin{figure*}
\centerline{
\includegraphics[width=\textwidth]{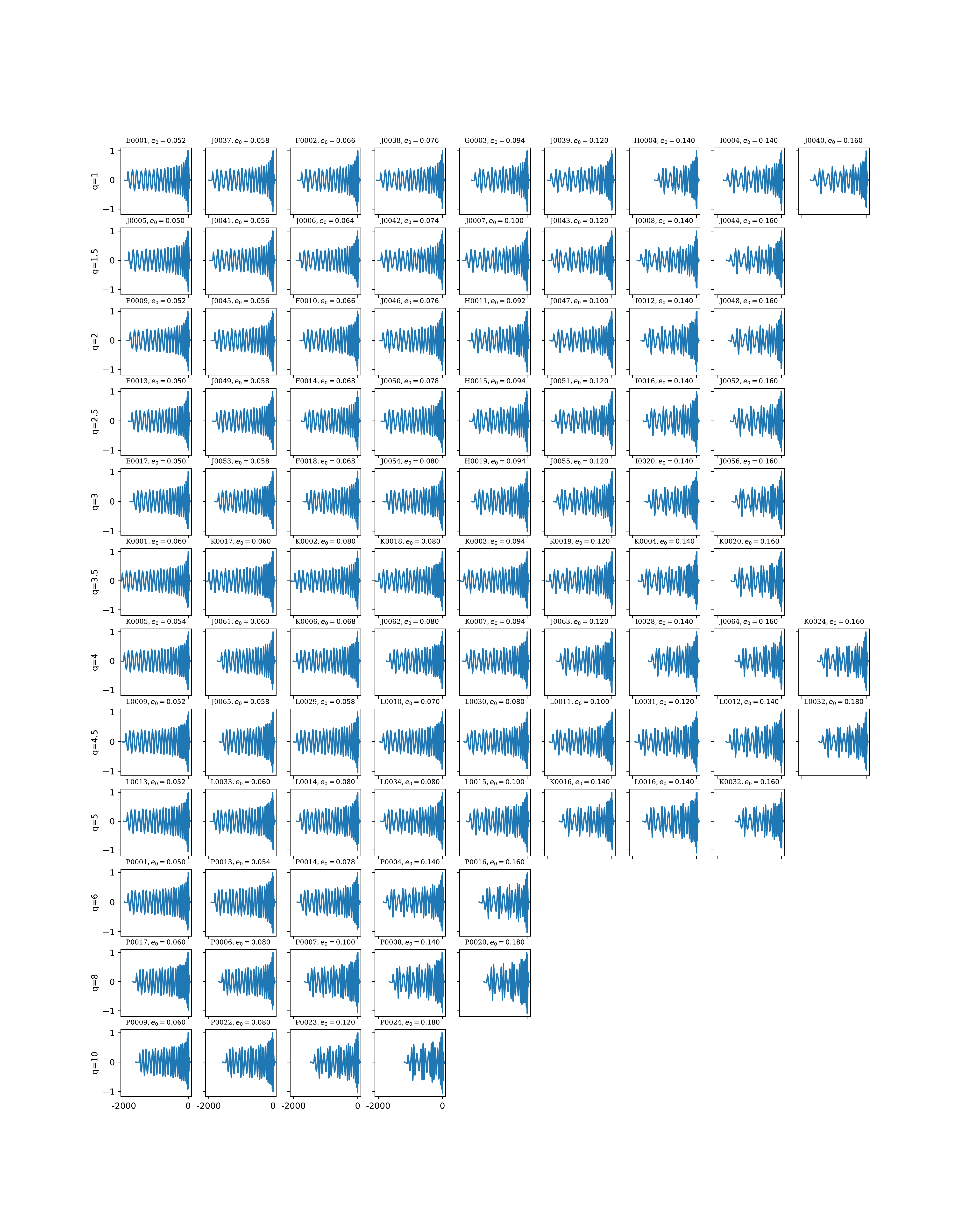}
}
\caption{For a given mass-ratio \(q\), each row presents the real part of the \(\ell=\abs{m}=2\) mode of each waveform in our catalog, extracted at future null infinity. The initial eccentricity, \(e_0\), increases from left to right. All these waveforms have unit amplitude at \(t=0\). Table~\ref{properties} lists the properties of this waveform catalog.} 
\label{catalog}
\end{figure*}

We have produced a catalog of 89 simulations with the open source, NR software, \texttt{the Einstein Toolkit}~\cite{etweb,Loffler:2011ay,naka:1987,shiba:1995,baum:1998,baker:2006,camp:2006,Lama:2011,wardell_barry_2016_155394,Pollney:2009yz,Thomas:2010aa,ETL:2012CQGra,Ansorg:2004ds,Diener:2005tn,Dreyer:2002mx,Schnetter:2003rb,Thornburg:2003sf,Brown:2008sb,Husa:2004ip,Kranc:web}. This catalog describes non-spinning BBHs with mass-ratios \(1\leq q \leq 10\) and eccentricities as high as \(e_0=0.18\) fifteen cycles  before merger. A visualization of this catalog may be found at~\cite{catalogI,catalogII}. We have post-processed the data products of these simulations using the open source software stack \texttt{POWER}~\cite{johnson:2017}, and extracted the modes \((\ell, \, \abs{m})= \{(2,\,2),\, (2,\,1),\, (3,\,3),\, (3,\,2), \, (3,\,1),\, (4,\,4),\, (4,\,3),\, (4,\,2)\), \((4,\,1)\}\). As described in Appendix~\ref{conv}, each of these simulations was produced with several levels of resolution to quantify convergence. The real part of the \(\ell=\abs{m}=2\) mode, extracted at future null infinity, for each NR waveform is presented in Figure~\ref{catalog}. The properties of these NR waveforms are listed in Table~\ref{properties}.

\noindent Characterizing the properties of the NR waveforms presented in Table~\ref{properties} requires the construction of 
a method to quantify the orbital eccentricity of these simulations. Using the orbital evolution of these simulations to obtain an estimate of the orbital eccentricity is inadequate due to the gauge-dependent nature of the binary's orbit.
On the other hand, methods to construct initial data for spinning BHs on quasi-circular orbits have also introduced definitions of orbital eccentricity, based on orbital separations and waveform phase and amplitude of the Weyl scalar \(\psi_4\)~\cite{initial_data_RIT:2017CQG}. However, while the scope of the method introduced in~\cite{initial_data_RIT:2017CQG} is to construct high-quality initial data for quasi-circular mergers, and therefore, using \({\cal{O}}(e)\) approximations to model the effect of eccentricity may suffice, we aim to measure larger values of orbital eccentricity.

 To address this matter, we have used the inspiral-merger-ringdown \texttt{ENIGMA} waveform model introduced in~\cite{huerta:2018PhRvD} to determine the eccentricity, mean anomaly and gauge-invariant frequency parameters, (\(e_0, \, \ell_0, \, x_0\)), that optimally describe each NR waveform in our catalog. We do this by finding the (\(e_0,\, \ell_0, \, x_0\)) triplet that maximizes the overlap between each NR waveform and its \texttt{ENIGMA} counterpart. In~\cite{habhu:2019} we quantified the optimal time window to remove junk radiation while keeping intact the signatures of eccentricity at early times in the NR waveforms. Such time range is given by \(t\leq60M\). A detailed description of this method, including the corresponding open source software stack for its use to characterize NR waveforms catalogs at scale, is presented in an accompanying article~\cite{habhu:2019}. 

In brief, we construct our method using the inspiral evolution of the \texttt{ENIGMA} waveform model, which contains state-of-the-art post-Newtonian corrections for eccentric binaries, which include eccentricity corrections in the conservative and radiative pieces up \({\cal{O}}(e^{12})\), including instantaneous, tails and tails-of-tails contributions, and a contribution due to
nonlinear memory; and quasi-circular corrections both from post-Newtonian, self-force and perturbative calculations up to \({\cal{O}}(x^{6})\)~\cite{Huerta:2017a,huerta:2018PhRvD}. Furthermore, in~\cite{huerta:2018PhRvD}, we have demonstrated 
that \(e_0=0\) \texttt{ENIGMA} waveforms capture the dynamics of quasi-circular BBH mergers 
with excellent accuracy. We showed this by computing overlaps between 
quasi-circular \texttt{ENIGMA} waveforms and their quasi-circular Effective One Body (EOB) counterparts~\cite{Bohe:2016gbl}. Assuming advanced LIGO's 
Zero Detuned High Power sensitivity~\cite{ZDHP:2018}, and using an initial GW frequency of 15 Hz to 
compute the overlaps, 
Figure 2 in~\cite{huerta:2018PhRvD} shows that the overlap between quasi-circular  \texttt{ENIGMA}  and EOB waveforms 
is \({\cal{O}}\geq 0.99\). Since the waveforms we are characterizing in this 
study are much shorter than those used to assess the accuracy of the \texttt{ENIGMA}  model in 
the quasi-circular limit, it follows that  \texttt{ENIGMA}  will capture the dynamics of moderately 
eccentric systems with excellent accuracy.

It is worth highlighting that while the \texttt{ENIGMA} waveform model was originally \textit{validated} with eccentric NR waveforms that describe BBH mergers with mass-ratios \(q\leq5.5\) and eccentricities \(e_0\leq0.18\) twenty cycles before merger~\cite{huerta:2018PhRvD}, it is through this analysis, and with the availability of new NR waveforms, that we can now report that the \texttt{ENIGMA} model can accurately describe BBH mergers with mass-ratios up to \(q=10\) with \(e_0 \leq 0.18\) fifteen cycles before merger.  

\section{Energy and angular momentum emission of eccentric black hole mergers}
\label{enbbh}

\begin{figure}[H]
\includegraphics[width=0.51\textwidth]{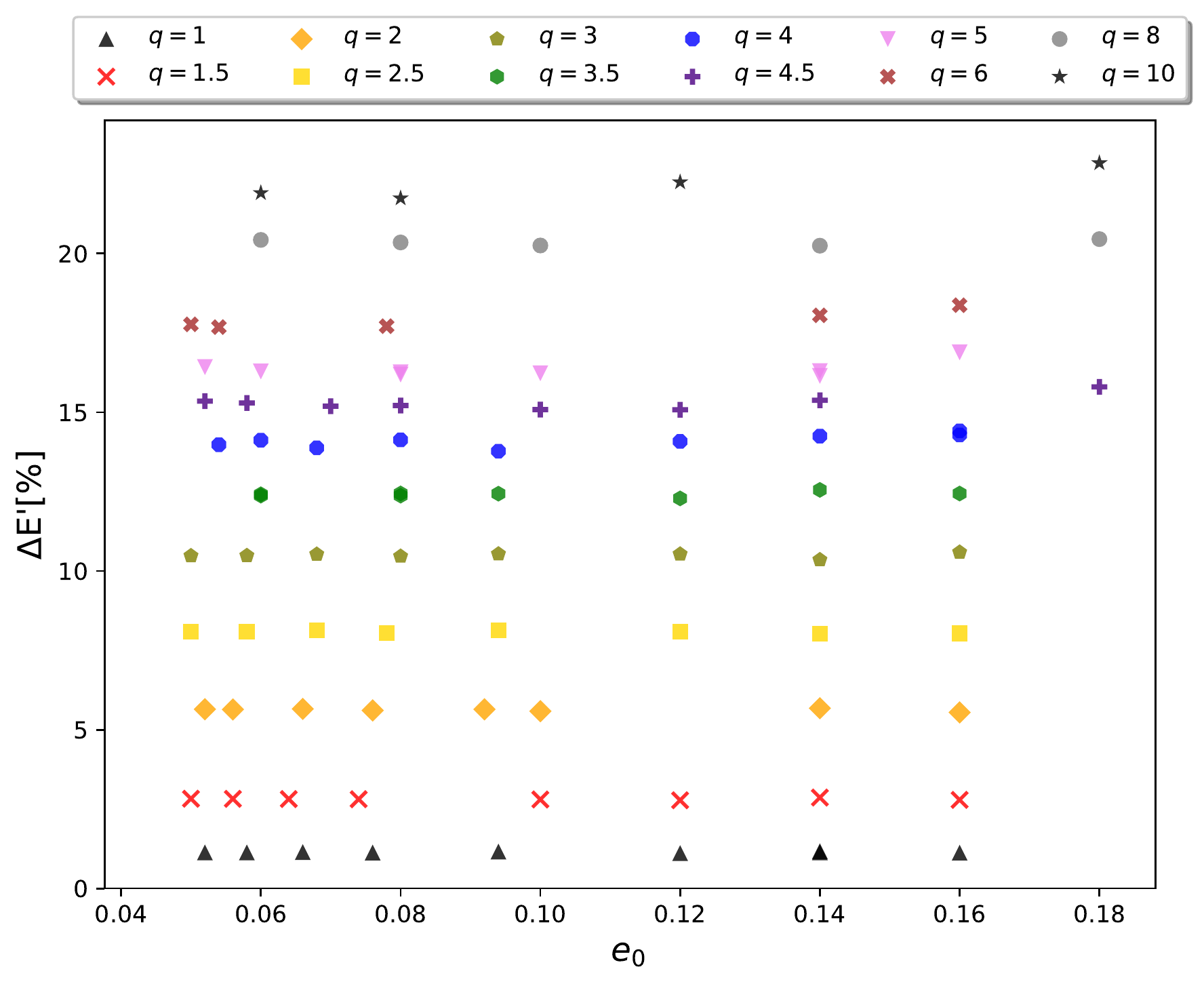}
\includegraphics[width=0.51\textwidth]{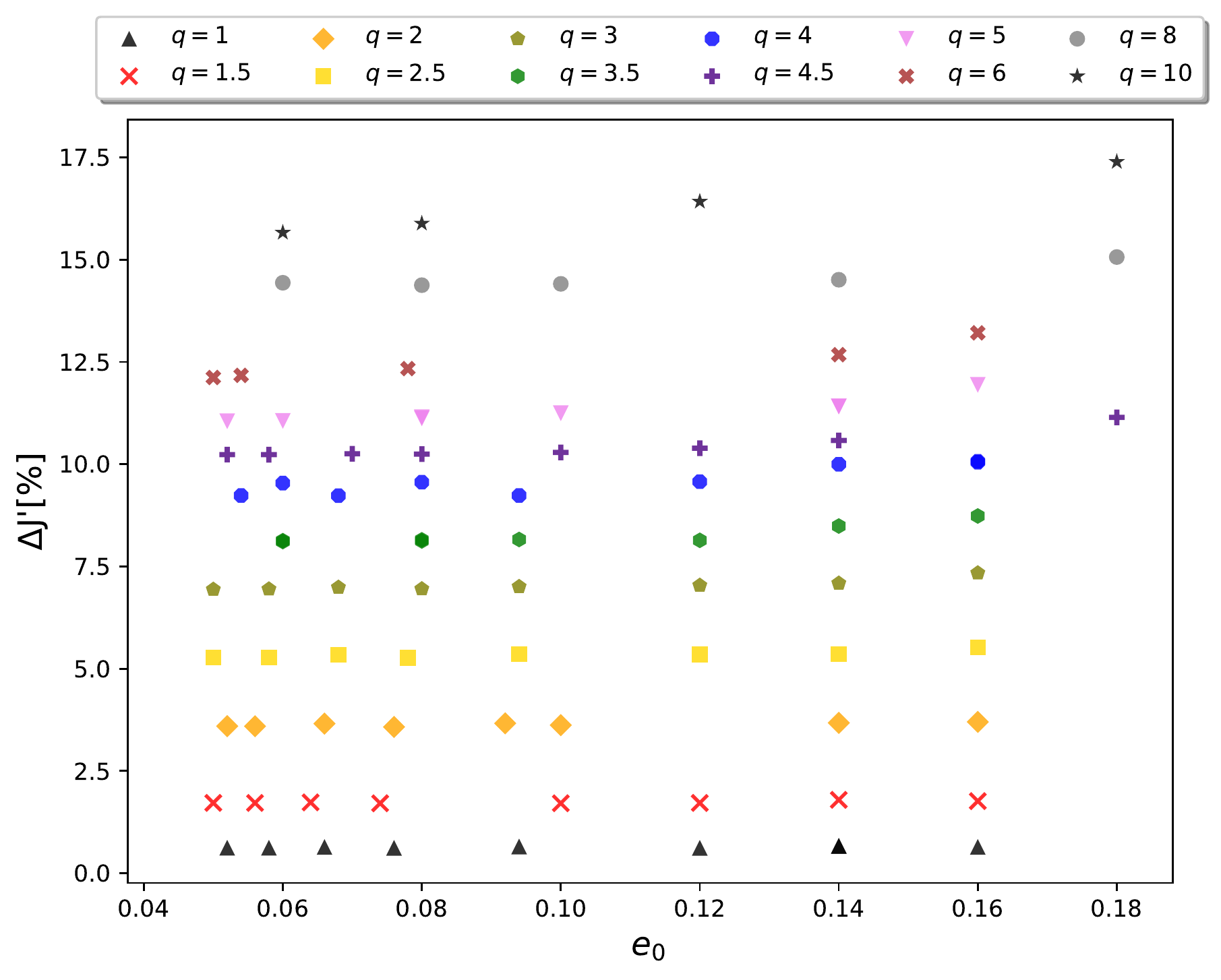}
\caption{Top panel: pair-wise comparison in radiated energy between NR waveforms that include either all \((\ell,\abs{m})\) modes or just the \(\ell=\abs{m}=2\) mode, as described by \(\Delta E'\) in Eq.~\eqref{e_diff}. Bottom panel: similar to the top panel, but now for radiated angular momentum, given by \(\Delta J'\) in Eq.~\eqref{j_diff}.}
\label{fig:energy}
\end{figure}

For each NR waveform in our catalog, we have quantified the energy, \(E\), and angular momentum, \(J\), radiated away through GW emission using the relations~\cite{damour:2012PhRvLD}

\begin{align}
\label{energy} 
\Delta E  & = \frac { 1 } { 16 \pi } \int _ { t _ { 0 } } ^ { t }\sum _ { m = - \ell } ^ { \ell } \sum _ { \ell = 2 } ^ { \ell _ { \max } }  \mathrm { d } t ^ { \prime } \left| N_{ \ell m } \left( t ^ { \prime } \right)\right|^2\,,\\
\Delta J  & = \frac { 1 } { 16 \pi } \int _ { t _ { 0 } } ^ { t }\sum _ { m = - \ell } ^ { \ell } \sum _ { \ell = 2 } ^ { \ell _ { \max } }  \mathrm { d } t ^ { \prime } m \Im \left[h_{\ell m}\left(t'\right) N^{*}_{ \ell m } \left( t ^ { \prime } \right)\right]\,, \\
\label{news}
N_{ \ell m } (t)& = \frac{\mathrm{d} h_{\ell m} (t)}{\mathrm{d}t}\,,
\end{align}

\noindent where \(N^{\ell m}(t)\) represents the complex \textit{news function} at infinity. The integration is done from the time the NR waveform is free from junk radiation, \(t_0=60M\), to the final sample time of the NR waveform, \(t\).  For these calculations we have considered the \((\ell, \, \abs{m})= \{(2,\,2),\, (2,\,1),\, (3,\,3),\, (3,\,2), \, (3,\,1),\, (4,\,4),\, (4,\,3),\, (4,\,2)\), \((4,\,1)\}\) modes. It is worth pointing out that the choice \(t_0\rightarrow 60M\) is informed by the study presented in~\cite{habhu:2019}, which demonstrated that this choice removes high-frequency noise while keeping intact the imprints of eccentricity in the NR waveforms once they are free from junk radiation. Using Eqs.~\eqref{energy}-\eqref{news}, in Figure~\ref{fig:energy} we quantify the importance of including higher-order waveform modes to compute the energy and angular momentum carried away by GWs. We do this through pair-wise comparisons between NR waveforms that include either all the modes listed above, or just the \(\ell=\abs{m}=2\) mode, using the relations 

\begin{eqnarray}
\label{e_diff}
\Delta E' &=& \frac{\Delta E(\ell, \abs{m}) - \Delta E(\ell=\abs{m}=2)}{\Delta E(\ell, \abs{m})}\,,\\
\label{j_diff}
\Delta J' &=& \frac{\Delta J(\ell, \abs{m}) - \Delta J(\ell=\abs{m}=2)}{\Delta J(\ell, \abs{m})}\,.
\label{diff}
\end{eqnarray}

\noindent Using the two highest resolution runs for each simulation in our catalog, 
we computed \((\Delta E',\, \Delta J')\), and found that the largest difference 
between these two independent measurements is \(\leq5\%\). The values we present in  Figure~\ref{fig:energy} were extracted from the highest resolution runs. We notice that for each mass-ratio BBH population, i.e., if we consider a given set of markers in the panels of Figure~\ref{fig:energy}, \((\Delta E',\, \Delta J')\)  are nearly constant across the eccentricity range that we have explored in this study. In different words, \((\Delta E',\, \Delta J')\) are constant polynomials in eccentricity for \(e_0\leq0.2\). We observe a minor deviation from this pattern at the high-end of the eccentricity range for the most asymmetric mass-ratio BBH systems. To be precise, if we fit a constant polynomial using the two lowest eccentricity samples for each mass-ratio population, we find that the largest deviation occurs for the most eccentric sample of the \(q=10\) BBHs, with a fractional error \(\leq8\%\) for the measurement of \(\Delta J' \).

\noindent These results also show that that for systems with \(q\geq5\) it is essential to include higher-order waveform modes to accurately describe the dynamics of eccentric BBH mergers. This result is consistent with recent studies~\cite{Adam:2018arXiv}, which indicate that the inclusion of higher-order modes for non-spinning, eccentric BBH mergers  has a more significant impact for GW detection, in the context of signal-to-noise ratio calculations, than for their non-spinning, quasi-circular BBH counterparts. It is worth highlighting that the eccentric NR waveform we have produced for this analysis for \(q\geq5\) are the first of their kind in the literature, so these results shed new light on the importance of including higher-order waveforms modes for the modeling of radiated energy and angular momentum of eccentric BBH mergers.

\section{Final mass, spin, and recoil velocity of post-merger black holes}
\label{fin}

We have computed the final mass, \(M_f\), and 
final spin, \(q_f\), of the BH remnant using the
\texttt{QuasiLocalMeasures} thorn of the \texttt{Einstein Toolkit}. The final mass is 
given by

\begin{align}
\label{final_mass}
M_f &= \sqrt{M_{\text{irr}}^2 + \frac{q_f^2}{4 M_{\text{irr}}}}\,,\quad \textrm{where}\\
\label{irrmass}
M_{\text{irr}} &= \sqrt{\frac{A}{16\pi}}\,,
\end{align}

\noindent \(M_{\text{irr}}\) is the irreducible mass, given in terms of the BHs' event horizon area, \(A\). 
$q_f$ is computed as the Komar angular momentum~\cite{mtw,WaldBook:1984,Poisson:2009pwt,Jaramillo:2010ay}

\begin{align}
\label{fspin}
q_f = \frac{1}{8\pi}\oint_S K_{ij} s^i \phi^j\,dA\,,
\end{align}

\noindent where the integral is over the surface, $S$, of the apparent horizon, $K_{ij}$
is the extrinsic curvature, $s^i$ is a spacelike, outward normal to the
horizon, and $\phi^i$ is a Killing vector associated with the rotational
symmetry around the spin axis.

\begin{figure}[H]
\includegraphics[width=0.51\textwidth]{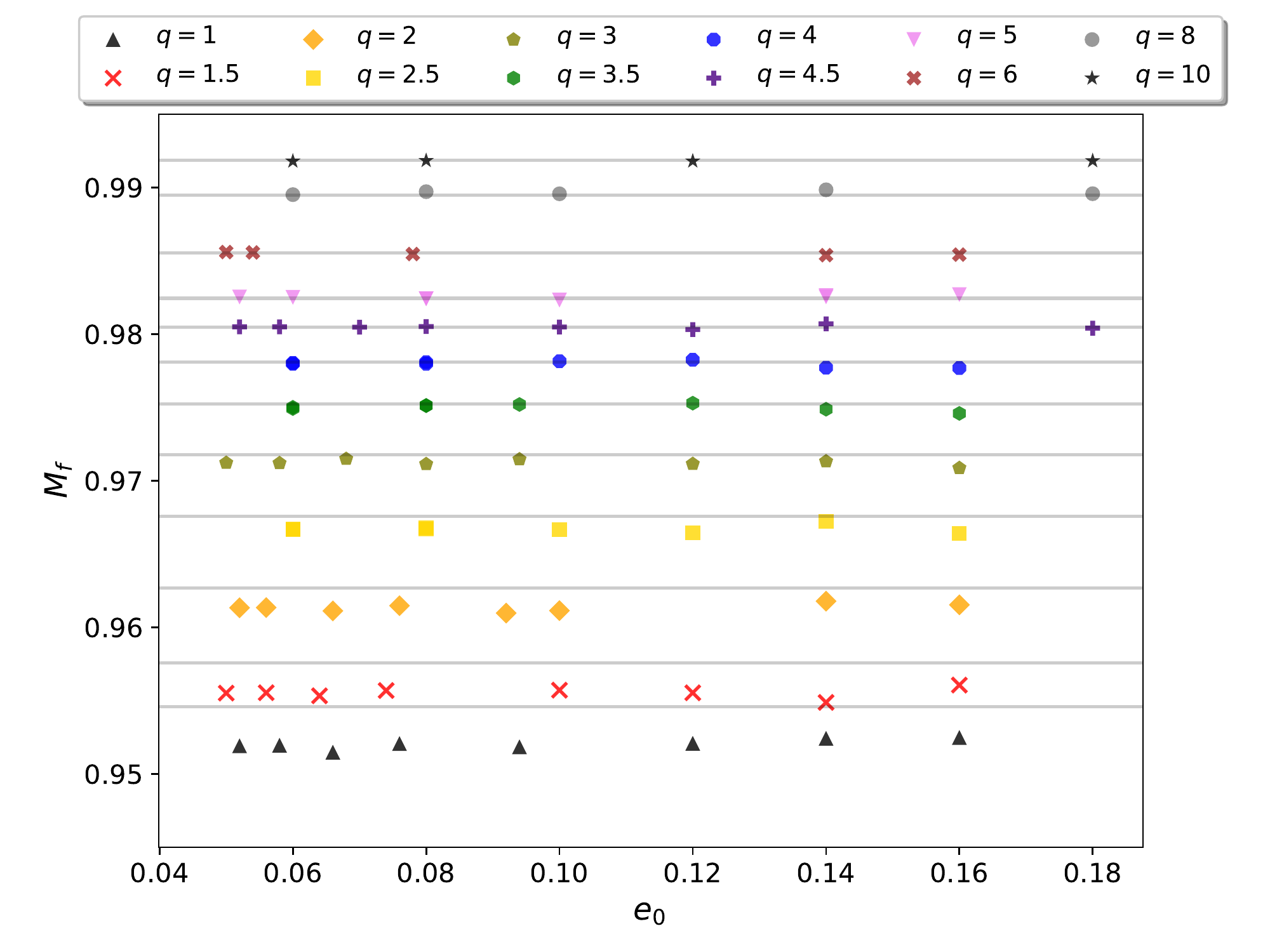}
\includegraphics[width=0.51\textwidth]{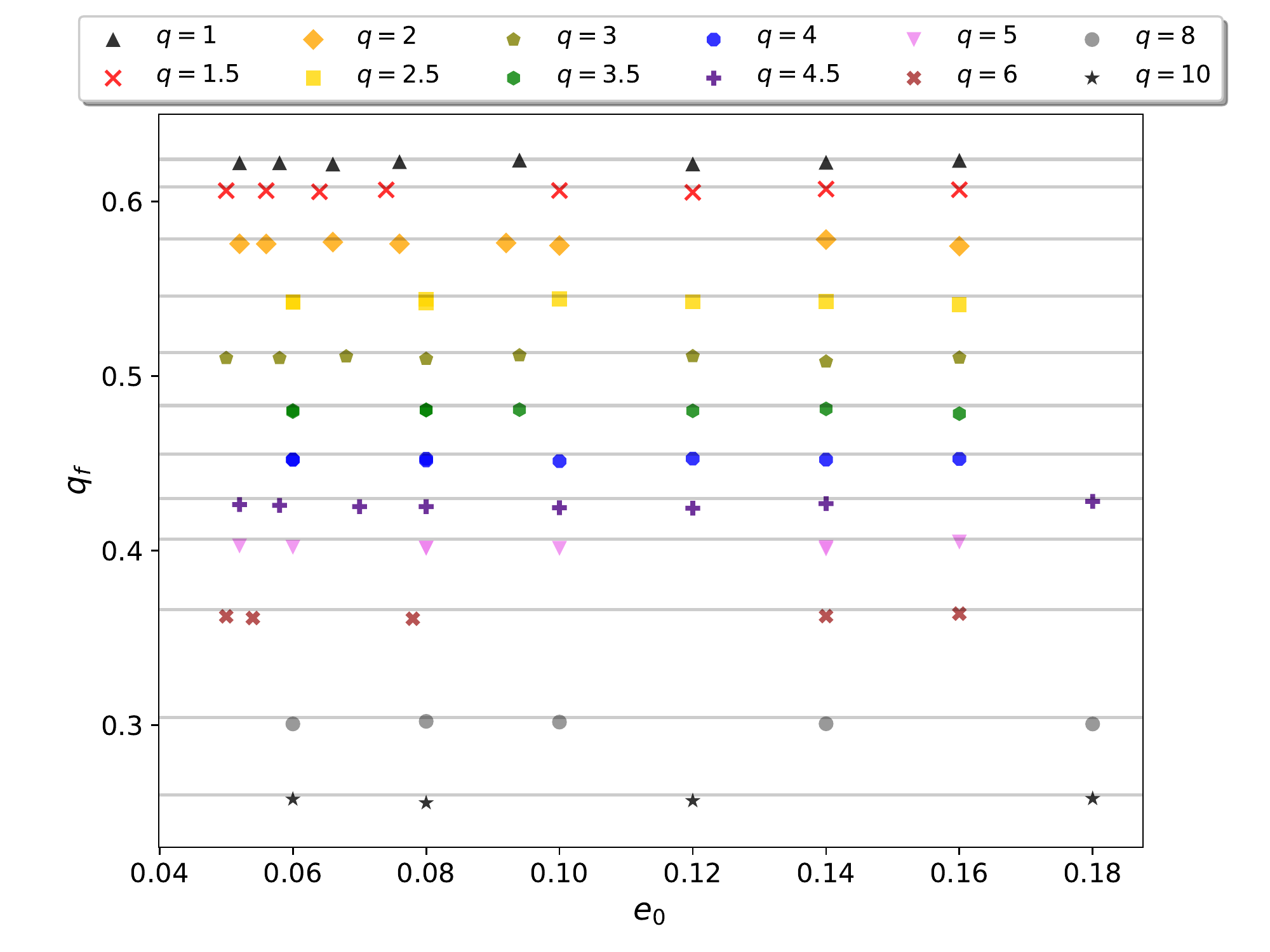}
\caption{Final mass, \(M_f\), (top panel) and final spin, \(q_f\), (bottom panel) of the black hole remnant as a function of the initial eccentricity, \(e_0\), and mass-ratio, \(q\) of the binary black hole systems listed in Table~\ref{properties}.} 
\label{massspin}
\end{figure}

In Figure~\ref{massspin} we present results for \((M_f,\,q_f)\). As before, 
we used our two highest resolution runs for each NR simulation to compute these observables, 
and found that these two independent measurements differ by \(\leq 3\%\). The results 
presented for \((M_f,\,q_f)\) in Figure~\ref{massspin} were extracted from the highest resolution 
runs in our catalog. We notice that for each mass-ratio population, i.e., for a given set of markers, the final mass and spin of the BH remnant are nearly independent of eccentricity in the range \(e_0\leq0.2\). We conclude this since both \((M_f,\,q_f)\) can be described as constant polynomials in eccentricity within the range we have considered in this study. We can directly compare these results using formulae derived for quasi-circular BBH mergers in~\cite{Baker:2008,hofman:2016ApJH}. Notice that we have included horizontal gray lines in both panels that provide the predictions for \((M_f,\,q_f)\) in the \(e_0\rightarrow0\) limit. 

\begin{figure}[h]
\includegraphics[width=0.48\textwidth]{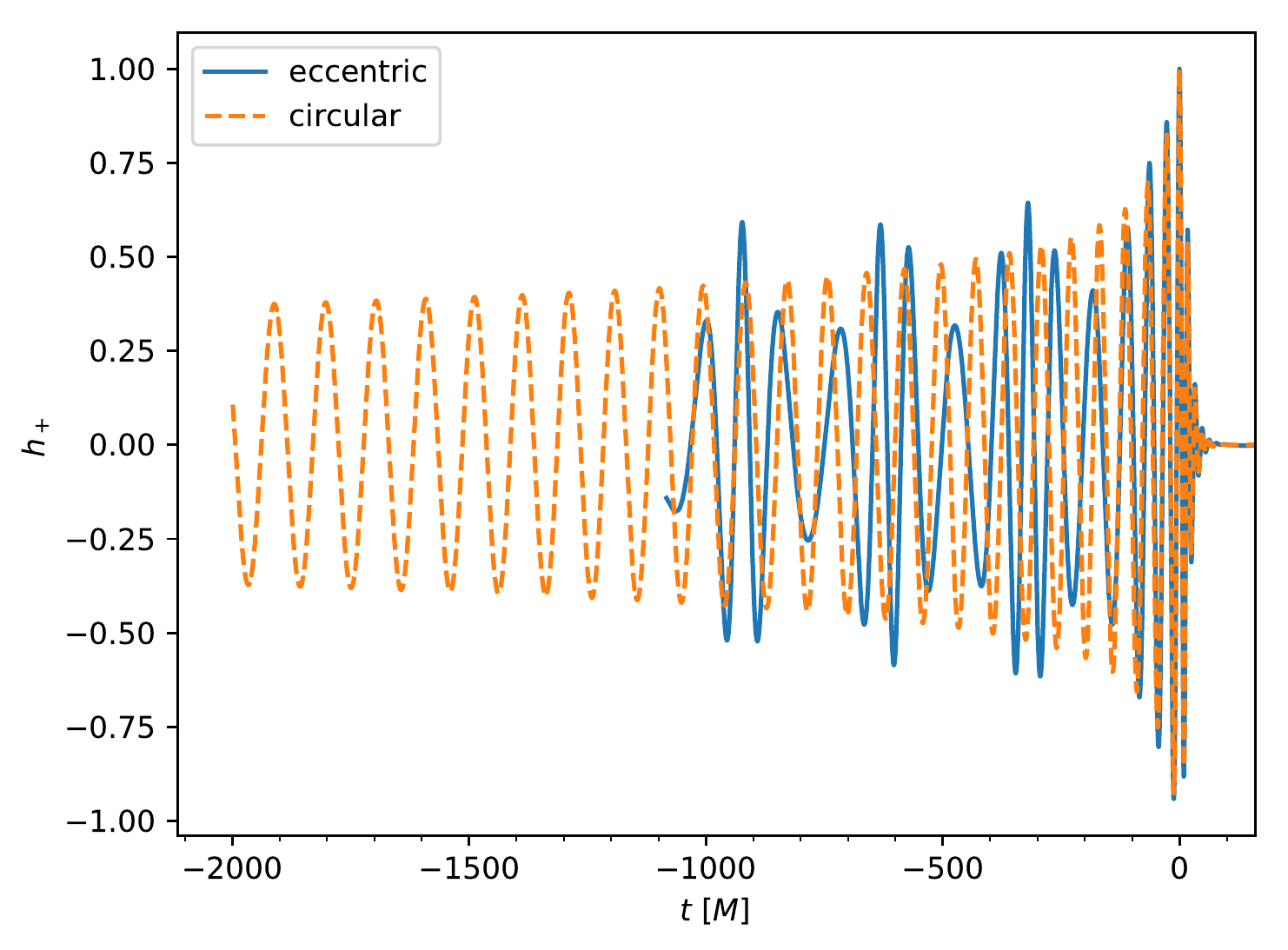}
\caption{Waveform signals produced by \(q=10\) BBHs that have the same orbital separation, but different initial eccentricity. We notice that even though the eccentric system has a large initial eccentricity, \(e_0=0.18\) fifteen cycles before merger, the waveform signal a few cycles before merger is consistent with a quasi-circular BBH system.} 
\label{circularize}
\end{figure}

For the \(M_f\) results (top panel in Figure~\ref{massspin}) the gray lines present the quasi-circular prediction for the final mass of the BH remnant for the mass-ratios \(q=\{1,\,1.5,\,2,\,2.5,\,3,\,3.5,\,4,\,4.5,\,5,\,6,\,8,\,10\}\) from bottom to top, respectively. We notice that the equal mass-eccentric BBH population presents the largest deviation from the quasi-circular prediction. However, this discrepancy is \(\leq1\%\). In the case of the final spin of the BH remnant, the bottom panel of Figure~\ref{massspin} also presents the quasi-circular predictions for this observable. Notice, however, that in this case, the grey lines describe the mass-ratios listed above but now from top to bottom. As in the case of \(M_f\), our results for the final spin of moderately eccentric BBH mergers are fairly consistent with results obtained from quasi-circular BBH mergers. This can only the case if the eccentric NR waveforms we have produced in this catalog circularize prior to merger. We have explored this scenario in detail, and have found that this is indeed the case. For a sample case, Figure~\ref{circularize} presents two waveform signals produced by BHs that have the same separation, but different initial eccentricity. The eccentric waveform contains all the telltale signatures of eccentricity, i.e., significant modulations in the amplitude and phase at early times, which correspond to periapse (local maxima) and apoapse (local minima) passages. We also observe that the waveform circularizes very rapidly, from \(e_0=0.18\) fifteen cycles before merger, turning into a quasi-circular waveform signal near the merger event. This is the reason why the results presented in Figure~\ref{massspin} are consistent with their quasi-circular counterparts. 

Earlier work on this front includes~\cite{ihh:2008PhRvD}, which presented calculations for the final spin and circularization of equal-mass eccentric BBH mergers, and showed that for BBH mergers with larger initial eccentricities than those considered in this work, the final spin of the BH remnant is greater than its quasi-circular counterpart. Additionally, reference~\cite{hinder:2017a} discussed the circularization of moderately eccentric BBH mergers with \(q\leq3\). In this article, we provide a systematic study of the observables \((q_f,\,M_f)\) to furnish evidence for the circularization of eccentric BBH mergers with \(q\leq 10\) and \(e_0\lesssim0.18\) fifteen cycles before merger. 

We have also computed the recoil velocity of eccentric BBH mergers,  \(\abs{v}_{\textrm{kick}}\), using the \texttt{PunctureTracker} thorn in the \texttt{Einstein Toolkit}. To do this, we have considered the last 100M of evolution of our NR simulations, and a simple first order finite difference formula for the velocity in terms on the measured locations. In Table~\ref{tab:recoil} we present the minima and maxima of the recoil velocity, \(\abs{v}_{\textrm{kick}}\), for the range of eccentricities we consider for each mass-ratio. We have also obtained gauge-invariant perturbations following~\cite{Pollney:2007ss} (see Eqs. (33)--(39) therein). These gauge-invariant results are presented in the last column of Table~\ref{tab:recoil}.

Important observations to be drawn from Table~\ref{tab:recoil} include: (i) the kick velocity of BH remnants produced by quasi-circular BH mergers, \(\abs{v}^{e_0\rightarrow0}_{\textrm{kick}}\), was obtained using the formulae presented in~\cite{recoil_2013PhRvDL};  (ii) the recoil velocity for all the \(q=1\) eccentric BBH mergers in our catalog is \(\abs{v}_{\textrm{kick}}=0\), which is consistent with results obtained for non-spinning, quasi-circular BBH mergers~\cite{koppitz_2007PhRvLK}; (iii) the kick velocities for our \(1\leq q\leq10\) population of eccentric BBH mergers are fairly consistent with the expected values of their quasi-circular counterparts, even though the formulae used to estimate \(\abs{v}^{e_0\rightarrow0}_{\textrm{kick}}\) was calibrated with quasi-circular BBH mergers with mass-ratios \(q\leq8\). 

\begin{table}[h!]
\setlength{\tabcolsep}{10pt}
  \begin{center}
    \caption{The table presents, from left to right, the mass-ratio of binary black hole mergers, \(q\), the recoil velocity of quasi-circular binary black hole mergers, \(\abs{v}^{e_0\rightarrow0}_{\textrm{kick}}\), and the \texttt{minimum} and \texttt{maximum} recoil velocities of our numerical relativity catalog for a given mass-ratio population, \([\abs{v}^{\texttt{min}}_{\textrm{kick}},\,\abs{v}^{\texttt{max}}_{\textrm{kick}}]\).}
    \label{tab:recoil}
    \begin{tabular}{c|c|c}
      \textbf{$q$} & $\abs{v}^{e_0\rightarrow0}_{\textrm{kick}}\, [\textrm{km}/\textrm{s}] $& $[\abs{v}^{\texttt{min}}_{\textrm{kick}},\,\abs{v}^{\texttt{max}}_{\textrm{kick}}]\, [\textrm{km}/\textrm{s}]$ \\[3pt]
      \hline\hline
      1.0 & 0.0   & [0.0,\,0.0]       [0.0,\,0.0]  \\
      1.5 & 107.4 & [94.6,\,101.8]    [104.0,\,109.9]\\
      2.0 & 156.7 & [136.9,\,149.0]   [115.1,\,157.5] \\
      2.5 & 173.5 & [152.8,\,165.0]   [132.9,\,180.7]\\
      3.0 & 174.1 & [161.4,\,170.9]   [151.8,\,179.1]\\
      3.5 & 167.1 & [154.4,\,173.7]   [144.0,\,178.8]\\
      4.0 & 156.9 & [143.1,\,166.8]   [140.0,\,173.8]\\
      4.5 & 145.6 & [137.9,\,154.6]   [133.6,\,157.1]\\
      5.0 & 134.4 & [121.0,\,137.1]   [117.6,\,136.8]\\
      6.0 & 113.9 & [99.5,\,121.0]    [104.5,\,121.3]\\
      8.0 &  82.7 & [88.7,\,96.5]     [70.6,\,90.4]  \\
     10.0 &  61.9 & [47.0,\,78.6]     [53.9,\,66.4]  \\
      \hline\hline
    \end{tabular}
  \end{center}
\end{table}

These results for \((M_f,\,q_f,\,\abs{v}_{\textrm{kick}})\) cover an entirely new region of parameter space in the modeling of eccentric BBH mergers, providing new insights into the physics of these GW sources. We discuss the implications of these findings in the following section.

\section{Implications for the modeling and detection of eccentric mergers}
\label{mol}

To date, there are only a handful of inspiral-merger-ringdown waveform models that describe 
the GW emission of eccentric BBH mergers~\cite{huerta:2018PhRvD,Huerta:2017a,hinder:2017a,Hinderer:2017,cao:2017}. These models assume that moderately eccentric BBHs circularize prior to the merger event. This assumption is sound, in light of the results presented in the previous section, for BBHs with \(q\leq10\) and whose residual eccentricity is as high as \(e_0\leq0.18\) just fifteen cycles before merger. 

Furthermore, we have found that for the most extreme sample of our NR catalog, e.g., P0024, which represents BBHs with \(q=10\) and \(e_0=0.18\) fifteen cycles before merger, circularization is only attained right before merger, as shown in Figure~\ref{circularize}. In different words, while assuming circularization of moderately eccentric BBH mergers is a reasonable \textit{ansatz}, this also means that the modeling of these GW sources demands the development of an inspiral evolution scheme that provides an accurate description of the dynamical evolution of these objects throughout the inspiral evolution, and which remains accurate one or two cycles before merger.  To accomplish this level of accuracy so late in the inspiral evolution, we showed in~\cite{huerta:2018PhRvD,Huerta:2017a} that the inspiral evolution should include, at the very least, higher-order eccentric post-Newtonian corrections for the instantaneous and tails and tails-of-tails pieces, as well as contributions due to non-linear memory, and higher-order self-force and BH perturbation theory corrections. 

Future source modeling efforts to describe the inspiral evolution of spinning BBHs on eccentric orbits should include new developments from post-Newtonian, self-force and perturbation theory formalisms~\cite{binidam:2016,binida:2016PhRvD,bini:2018PhRvD,binidam:2018PhRvD,kavbini:2017PhRvD,binithi:2016PhRvD,Hinderer:2017,akcavan:2016PhRvD,ackay:2017CQG,fujite_2017CQGF,lt_2015PhRvDL,osburn_for_2014PhRvD,war_2017_PhRvDW,vdm_2016PhRvDV,vdm_2017PhRvL1V,vdm_2018PhRvD_97V,war_vdm_2018CQG3V,Ireland:2019tao}. These schemes may be complemented with stand-alone merger models designed with machine learning, or by directly attaching merger waveforms from NR surrogate waveform families~\cite{blackman:2015,varma:2018V,doctor:70605408D,Moore160125,huerta:2018PhRvD,gpr:2016PhRvD,Varma:2019csw}. The validation of these models with eccentric NR simulations will be essential to assess their accuracy and reliability for the detection and characterization of compact binary populations in dense stellar environments.

This waveform catalog may also be used to assess the sensitivity of burst searches to detect eccentric BBH mergers~\cite{SKlimenko:2004CQGra,Klimenko:2004CQG,Sergey:2008CQG,Sergey:2016,Tiwari:2016}, and to train neural network models to detect and characterize these GW sources~\cite{Adam:2018arXiv,geodf:2017a,geodf:2017b}. These studies will be pursued using this NR waveform catalog.

\section{Conclusion}
\label{end}
 
 We have studied the physics of eccentric BBH mergers using a 
 NR waveform catalog that describes BBH systems with mass-ratios \(q\leq10\), 
 and initial eccentricities 
 \(e\leq 0.18\) up to fifteen cycles before merger. 
 
We quantified the importance of including higher-order 
waveform modes to compute the energy and angular momentum carried away by 
GWs in eccentric BBH mergers. 
We have also demonstrated that the 
 properties of BH remnants described by our NR catalog are consistent with their quasi-circular 
 counterparts, which provides evidence for the circularization of moderately eccentric BBH mergers. We have also 
 computed recoil velocities of BH remnants produced by eccentric BBH mergers, and found that 
 these are fairly similar to those computed in the literature for non-spinning, quasi-circular BBH mergers.
 
 Based on these analyses, we have provided evidence that existing source modeling efforts that assume the 
 circularization of moderately eccentric BBH mergers is sound. We have also shown that since circularization 
 takes place close to the merger event, any semi-analytical model that is used to describe these GW sources should 
 include higher-order corrections both to the conservative and radiative pieces of the source's dynamics, and 
 to the waveform strain.

Recent studies in the literature have identified parameter space degeneracies between orbital eccentricity 
and spin corrections~\cite{Huerta:2017a}. In order to get better insights into this finding, it is essential to 
understand the dynamics of spinning BHs on eccentric orbits, and then use the NR waveform catalog we have introduced 
in this study to carefully assess in which regions of parameter space such a degeneracy may be broken to distinguish 
these two compact binary populations. 

The construction of a NR waveform catalog for spinning BHs on eccentric orbits is already underway to shed light on this timely, and astrophysically motivated study. Specific aspects to address in such study will encompass: (i) orbital configurations that significantly shorten the length of waveform signals, i.e., eccentricity and spin anti-aligned configurations; (ii) competing effects to determine the length of waveform signals, i.e., rapidly spinning BHs on spin-aligned configurations (which increase the length of waveforms as compared to non-spinning BBHs) vs moderate values of initial eccentricity (which decrease the length of waveforms as compared to quasi-circular BBHs); and (iii) identify telltale signatures of GW sources that can be used to infer the existence of eccentric compact binary populations through GW observations, e.g., astrophysical properties of the BH remnant, and the coupling of eccentricity and spin-spin and spin-orbit effects at peri-apse passages during the inspiral evolution of these systems. 

The modes we have extracted in this study with the open source \texttt{POWER}~\cite{johnson:2017} package do not include \(m=0\) memory modes. Their extraction requires the use of the Cauchy Characteristic Extraction method~\cite{Reisswig:2009rx}. Given the importance of these modes for the characterization of eccentric BBH mergers, we will present these modes in a forthcoming study, accompanied by a systematic analysis on the observability of these modes with second and third generation GW detectors.

\acknowledgments
This research is part of the Blue Waters sustained-petascale computing project, which is supported by the National Science Foundation (awards OCI-0725070 and ACI-1238993) and the State of Illinois. Blue Waters is a joint effort of the University of Illinois at Urbana-Champaign and its National Center for Supercomputing Applications. We acknowledge support from the NCSA and the SPIN Program at NCSA. We thank the \href{http://gravity.ncsa.illinois.edu}{NCSA Gravity Group} for useful feedback. NSF-1550514, NSF-1659702 and TG-PHY160053 grants are gratefully acknowledged.

\bibliography{references}
\bibliographystyle{apsrev4-1}

\appendix
\section{Properties of Numerical Relativity Catalog}

Table~\ref{properties} lists the properties of our numerical relativity catalog.

\begin{center}
\setlength{\tabcolsep}{10pt}
\setlength{\LTcapwidth}{3in}
\begin{longtable}{c c c c c}
\caption{$(e_0,\, \ell_0,\, x_0)$ represent the measured values of eccentricity, mean anomaly, and dimensionless orbital frequency parameters. These quantities are computed upon removing the first \(60M\) of evolution of the numerical relativity waveforms, as described in~\cite{habhu:2019}.}\\
\label{properties}\\
\hline 
\hline 
 Simulation &	$q$	&	$e_0$ & $\ell_0$ & $x_0$ \\
\hline
\hline
E0001	&	1	&	0.052	&	3.0	&	0.0770	\\
E0009	&	2	&	0.052	&	3.0	&	0.0794	\\
E0013	&	2.5	&	0.050	&	3.0	&	0.0813	\\
E0017	&	3	&	0.050	&	3.0	&	0.0831	\\
F0002	&	1	&	0.066	&	3.0	&	0.0780	\\
F0010	&	2	&	0.066	&	3.0	&	0.0803	\\
F0014	&	2.5	&	0.068	&	3.0	&	0.0822	\\
F0018	&	3	&	0.068	&	3.0	&	0.0842	\\
G0003	&	1	&	0.094	&	3.0	&	0.0788	\\
H0004	&	1	&	0.140	&	3.0	&	0.0826	\\
H0011	&	2	&	0.092	&	3.0	&	0.0795	\\
H0015	&	2.5	&	0.094	&	3.0	&	0.0812	\\
H0019	&	3	&	0.094	&	3.0	&	0.0832	\\
I0004	&	1	&	0.140	&	3.0	&	0.0765	\\
I0012	&	2	&	0.140	&	3.0	&	0.0791	\\
I0016	&	2.5	&	0.140	&	3.0	&	0.0811	\\
I0020	&	3	&	0.140	&	3.0	&	0.0824	\\
I0028	&	4	&	0.140	&	2.9	&	0.0865	\\
J0005	&	1.5	&	0.050	&	3.0	&	0.0779	\\
J0006	&	1.5	&	0.064	&	3.0	&	0.0782	\\
J0007	&	1.5	&	0.100	&	3.1	&	0.0762	\\
J0008	&	1.5	&	0.140	&	3.0	&	0.0768	\\
J0037	&	1	&	0.058	&	3.0	&	0.0768	\\
J0038	&	1	&	0.076	&	3.0	&	0.0762	\\
J0039	&	1	&	0.120	&	3.1	&	0.0749	\\
J0040	&	1	&	0.160	&	3.0	&	0.0761	\\
J0041	&	1.5	&	0.056	&	3.0	&	0.0777	\\
J0042	&	1.5	&	0.074	&	3.0	&	0.0771	\\
J0043	&	1.5	&	0.120	&	3.1	&	0.0756	\\
J0044	&	1.5	&	0.160	&	2.9	&	0.0778	\\
J0045	&	2	&	0.056	&	3.0	&	0.0793	\\
J0046	&	2	&	0.076	&	3.0	&	0.0787	\\
J0047	&	2	&	0.100	&	3.0	&	0.0778	\\
J0048	&	2	&	0.160	&	2.9	&	0.0794	\\
J0049	&	2.5	&	0.058	&	3.0	&	0.0811	\\
J0050	&	2.5	&	0.078	&	3.0	&	0.0806	\\
J0051	&	2.5	&	0.120	&	3.0	&	0.0795	\\
J0052	&	2.5	&	0.160	&	2.9	&	0.0817	\\
J0053	&	3	&	0.058	&	3.0	&	0.0829	\\
J0054	&	3	&	0.080	&	3.0	&	0.0823	\\
J0055	&	3	&	0.120	&	3.0	&	0.0816	\\
J0056	&	3	&	0.160	&	2.9	&	0.0829	\\
J0061	&	4	&	0.060	&	3.0	&	0.0855	\\
J0062	&	4	&	0.080	&	3.1	&	0.0847	\\
J0063	&	4	&	0.120	&	3.0	&	0.0841	\\
J0064	&	4	&	0.160	&	2.9	&	0.0863	\\
J0065	&	4.5	&	0.058	&	3.0	&	0.0878	\\
J0066	&	4.5	&	0.080	&	3.0	&	0.0870	\\
J0067	&	4.5	&	0.120	&	3.0	&	0.0858	\\
J0068	&	4.5	&	0.180	&	2.9	&	0.0874	\\
K0001	&	3.5	&	0.060	&	3.0	&	0.0802	\\
K0002	&	3.5	&	0.080	&	3.0	&	0.0808	\\
K0003	&	3.5	&	0.094	&	3.1	&	0.0800	\\
K0004	&	3.5	&	0.140	&	3.0	&	0.0810	\\
K0005	&	4	&	0.054	&	3.0	&	0.0817	\\
K0006	&	4	&	0.068	&	3.0	&	0.0826	\\
K0007	&	4	&	0.094	&	3.0	&	0.0823	\\
K0008	&	4	&	0.140	&	2.9	&	0.0833	\\
K0016	&	5	&	0.140	&	2.9	&	0.0868	\\
K0017	&	3.5	&	0.060	&	3.0	&	0.0801	\\
K0018	&	3.5	&	0.080	&	3.1	&	0.0801	\\
K0019	&	3.5	&	0.120	&	3.1	&	0.0789	\\
K0020	&	3.5	&	0.160	&	2.9	&	0.0829	\\
K0021	&	4	&	0.060	&	3.0	&	0.0821	\\
K0022	&	4	&	0.080	&	3.0	&	0.0823	\\
K0023	&	4	&	0.120	&	3.0	&	0.0817	\\
K0024	&	4	&	0.160	&	2.9	&	0.0856	\\
K0032	&	5	&	0.160	&	2.8	&	0.0888	\\
L0009	&	4.5	&	0.052	&	3.0	&	0.0839	\\
L0010	&	4.5	&	0.070	&	3.0	&	0.0841	\\
L0011	&	4.5	&	0.100	&	3.0	&	0.0837	\\
L0012	&	4.5	&	0.140	&	2.9	&	0.0849	\\
L0013	&	5	&	0.052	&	3.0	&	0.0854	\\
L0014	&	5	&	0.080	&	3.0	&	0.0856	\\
L0015	&	5	&	0.100	&	3.0	&	0.0853	\\
L0016	&	5	&	0.140	&	2.9	&	0.0862	\\
L0017	&	5.5	&	0.060	&	3.0	&	0.0869	\\
L0018	&	5.5	&	0.068	&	3.0	&	0.0878	\\
L0019	&	5.5	&	0.100	&	3.0	&	0.0869	\\
L0020	&	5.5	&	0.140	&	2.9	&	0.0882	\\
L0029	&	4.5	&	0.058	&	3.0	&	0.0844	\\
L0030	&	4.5	&	0.080	&	3.1	&	0.0835	\\
L0031	&	4.5	&	0.120	&	3.1	&	0.0827	\\
L0032	&	4.5	&	0.180	&	3.0	&	0.0849	\\
L0033	&	5	&	0.060	&	3.0	&	0.0852	\\
L0034	&	5	&	0.080	&	3.0	&	0.0852	\\
L0037	&	5.5	&	0.060	&	3.0	&	0.0870	\\
L0038	&	5.5	&	0.080	&	3.0	&	0.0870	\\
L0039	&	5.5	&	0.120	&	2.9	&	0.0867	\\
L0040	&	5.5	&	0.180	&	2.9	&	0.0894	\\
P0001	&	6	&	0.050	&	3.0	&	0.0867	\\
P0004	&	6	&	0.140	&	2.9	&	0.0867	\\
P0006	&	8	&	0.080	&	2.9	&	0.0931	\\
P0007	&	8	&	0.100	&	2.9	&	0.0926	\\
P0008	&	8	&	0.140	&	2.9	&	0.0910	\\
P0009	&	10	&	0.060	&	2.9	&	0.0971	\\
P0013	&	6	&	0.054	&	3.0	&	0.0871	\\
P0014	&	6	&	0.078	&	2.9	&	0.0885	\\
P0016	&	6	&	0.160	&	2.8	&	0.0900	\\
P0017	&	8	&	0.060	&	3.0	&	0.0927	\\
P0020	&	8	&	0.180	&	2.9	&	0.0936	\\
P0022	&	10	&	0.080	&	2.9	&	0.0979	\\
P0023	&	10	&	0.120	&	2.9	&	0.0968	\\
P0024	&	10	&	0.180	&	3.0	&	0.0957	\\
 \end{longtable}
 \end{center}
 
\section{Convergence of the numerical waveforms}
\label{conv}

We use a grid setup based on the setup used
in~\cite{wardell_barry_2016_155394}. There is a central, mesh refined cubical
region of the grid in which Cartesian coordinates are used, surround by 6
regions that make up a cubed sphere grid with constant angular resolution.

We use 8\textsuperscript{th} order finite differencing operators to compute
spatial derivatives of the spacetime quantities in the Einstein field
equations. This requires the use of $5$ ghost zones, and together with using a
classical 4\textsuperscript{4th} order Runge-Kutta timestepper implies that
each refined region is surrounded by $20$ points that are filled in via
prolongation from the next coarser region. We use vertex centered
5\textsuperscript{th} order prolongation operators rather than full
8\textsuperscript{th} order prolongation operators.

The cubical region employs mesh refinement with the resolution on the coarsest
grid being $h_{\text{coarse}} = 1.92\,M$. Each of the black holes is
surrounded by a set of nested moving boxes such that the resolution in the
finest box containing the black hole $i$ is $1.2 M_i / (N_l-1)$ where $M_i$ is
the initial mass parameter of black hole $i$ and $N_l$ is the number of points
used for the resolution level $l$ simulation. In our simulations we used $N_l
= 32, 36, 40, 44$, where $N_l = 44$ was only used for simulations with a mass
ratio $q > 5$. The finest box surrounding each black hole has a radius of
$1.2\,M_i$ and each coarser box has twice the radius of the next finer one.
During the simulation we track the location of each black hole and keep the
set of nested refined boxes approximately centered on the black hole.  Finally
the outer edge of the cubical region is chosen large enough to contain all
refined regions including their prolongation regions.

In the spherical region we choose an angular resolution of
$h_\text{angular} = \pi/\left(4\,N_l\right)$ and a radial resolution of $1.92\,M$
which matches the coarsest resolution in the Cartesian grid. The outer
boundary is chosen such that it is causally disconnected from the outermost
detector at which we extract gravitational waves from.

We use a time step $\Delta t = 0.864\,M$ on the coarsest level,
corresponding to a Courant-Friedrichs-Lewy condition of $\Delta t /
h_\text{coarse} = 0.45$ which is held constant on the finer levels by
decreasing their time step size.

We extract gravitational waves using modes of the Weyl scalar $\psi_4$
extracted on coordinate spheres of radius $r_{\text{det},i} = 100\,M, 115\,M,
136\,M, 167\,M, 214\,M, 300\,M,$ $500\,M$.

Using 8\textsuperscript{th} order
finite differencing operators our simulations would,
under ideal circumstances, converge towards the correct solution with an
error term which scales like $h^8$, where $h$ is the spatial resolution
of the simulation. However due to lower order schemes present in the
simulation, for example the interpolation at mesh refinement boundaries
which is only 5\textsuperscript{th} order accurate, as well as artifacts
caused by the adaptive mesh refinement logic making independent
decisions where to refine for each simulation, the observed convergence
order typically differs from 8.

\begin{figure}
\centering
\includegraphics[width=0.99\linewidth]{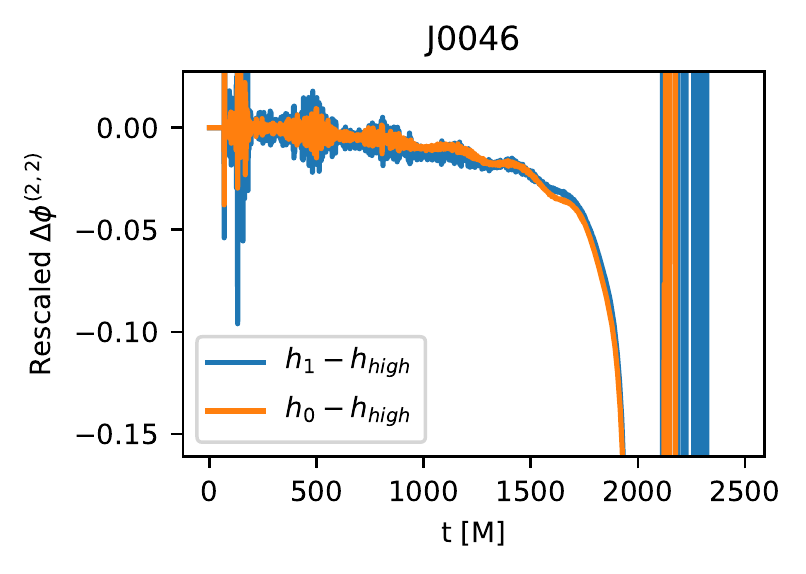}
\caption{Convergence of the phase difference in the gravitational wave
phase for case \texttt{J0046}, rescaled to demonstrate convergence at
order $N \approx 4.5$. We compute convergence in the time interval
$500\,M \le t \le t_{\text{max}} - 200\,M$, where $t_{\text{max}}$ is the
point of maximum amplitude, approximately corresponding to the time of
merger. Note that this plot includes the pulse of junk radiation due to our
initial data not containing any waves as well as the ringdown and merger
signal, both of which are not convergent and thus lead to very large phase
differences which are clipped in the plot.}\label{fig:convergence_J0046}
\end{figure}

To estimate the convergence of each waveform we simulated each set of
physical parameters using at least 3 (4) simulations using increasing
resolution for waveforms of mass ratio $q \le 5$ ($q > 5$). We then
compute the gravitational wave phase $\phi^{(2,2)}$ from the complex
$\ell = m = 2$ mode of the spherical harmonic decomposition of the
outgoing component of the Weyl scalar $\psi_4$ and studied its
convergence properties.

Figure~\ref{fig:convergence_J0046} shows the rescaled phase differences
$\phi^{(2,2)}(h_n) - \phi^{(2,2)}(h_{\text{high}})$ between the
gravitational wave phase obtained from the simulation with resolution
$h_n$ and the highest resolved simulation. Phase differences have been
rescaled such that for a convergent simulation the plotted curves
overlap. For case \texttt{J0046} we observe an approximate convergence
order of $N \approx 4.5$ which is within the range of expected values.

Not all simulated cases show clean convergence behaviour, with the
convergence order for some of them being larger than 8, which
may indicate that our lowest resolution simulation does not adequately
resolve the features present in the simulation domain, and others
swapping the ordering of phases between the low, medium and high
resolution simulations, making an estimate of the convergence order
impossible.

Given that large number of simulations, this is to be expected and does
not necessarily indicate that the obtained results are incorrect but
instead demonstrates the difficulty in controlling the various effects
that influence the numerically obtained waveform. Since there are
multiple sources of numerical error, and we have chosen parameters such
that none is dominant so as to make best use of available computing
resources without over-resolving a particular feature, different sources
of numerical error potentially cancel each other out, giving rise to
unrealistically large (or small) convergence orders.

\end{document}